\documentclass[a4paper]{article}
\usepackage[pdftex]{graphicx}
\usepackage{amsmath}
\usepackage{amssymb}
\usepackage{amsthm}
\usepackage[utf8]{inputenc}
\usepackage[T1]{fontenc}
\usepackage{hyperref}
\usepackage{smartref}
\usepackage{todonotes}
\usepackage{fullpage}
\usepackage{epstopdf}
\usepackage{subcaption}
\usepackage[inline]{enumitem}

\newtheorem{prop}{Proposition}
\newtheorem{lemma}{Lemma}

\newtheorem{corollary}{Corollary}
\usepackage[toc,page]{appendix}

\usepackage{color}

\renewcommand\vec{\boldsymbol}

\newcommand{\scal}[2]{\langle #1,#2\rangle}

\newcount\colveccount
\newcommand*\colvec[1]{
        \global\colveccount#1
        \begin{pmatrix}
        \colvecnext
}
\def\colvecnext#1{
        #1
        \global\advance\colveccount-1
        \ifnum\colveccount>0
                \\
                \expandafter\colvecnext
        \else
                \end{pmatrix}
        \fi
}

\begin{document}

\title{Projective Elasticity}
\author{Tam\'as Baranyai\footnote{Supported by the Ministry of Innovation
and Technology of Hungary from the National Research, Development and Innovation Fund under the
PD 142720 funding scheme.}}
\maketitle

\abstract
We present the foundations of a projective geometric theory of elasticity, as well as outline a few possible application possibilities. We give the description of the Cauchy stress and infinitesimal strain tensors compatible with coordinate description of projective geometry and derive their transformation rules under projective transformations. We identify these tensors with generalized conics and quadrics; which shows that the material law cannot be directly given as a correlation, and that the absence of normal stresses and normal strains are projective invariants. The transformation of the graph of the Airy stress function is given in a point-wise projective three dimensional way, which preserves Pucher's equation corresponding to a unidirectionally loaded membrane shell. The paper closes with an example considering tapered beams, showing how to find a statically possible stress distribution in trapezioidal plates relying on the Euler-Bernoulli beam theory for rectangular beams.

\section{Introduction}
Statics and instantaneous kinematics lend themselves naturally to a projective geometric treatment. This is known since Plücker \cite{plucker1868neue}, who, apart from introducing the projective line coordinates named after him gave the connection of mechanics with what he called right lines \cite{plucker_mechanics}. This connection is actively used in fields where one has to deal with spatial kinematics \cite{davidson2004robots,gallardo2016kinematic}, also relying on the work 
and terminology of Ball \cite{ball1900treatise}. This is not the case in civil engineering, at the time of writing this paper this geometric approach seems to have mostly died out from the common knowledge of civil engineers, in spite of attempts \cite{crapo1982statics} to revitalize it. This work argues for the projective treatment by extending the description of statics and instantaneous kinematics of rigid bodies to small-displacement elasticity. As in elasticity finding an analytical solution is not at all as simple as in statics, we expect the projective treatment to be even more useful. This is due to the hierarchy of transformations (projective being more general then affine) and an old idea to analyse structures through transformations \cite{rankine1857transformations}. The paper is structured as follows.\\

After recalling the necessary concepts from multi-linear algebra we give an algebraic description of projective elasticity. The stress and strain tensors are introduced in matrix form compatible with the matrix description of projective transformations. The transformation rules of the stress and strain tensors as well as those of body forces are worked out such that these transformations preserve static equilibrium and kinematic compatibility.  

As a geometric interpretation of the algebraic results generalized conics and quadrics are introduced corresponding to the tensors. This shows that the absence of normal stresses and normal strains are incidence conditions and thus are projective invariants. Together with the kernels of the projective tensors corresponding to subspaces this implies that projective transformations will naturally keep the static boundary conditions of the free edges and of tangential loads. It is also shown through this geometric treatment, that the material properties can not be represented through the duality present in projective geometry (at least not directly) and consequently we do not give an explicit transformation rule for the material law. The transformation laws of stress and strain tensors do implicitly imply the transformation law of the material properties.

Then the transformation of the graph of the Airy stress function is given in a point-wise projective three dimensional way. Through the use of Pucher's equation this allows for the description of transformations of membrane shells where the transformation is determined by what happens in the floor-plan of the shell. This way the mechanical behaviour of some skewed membrane shells may be to be reduced to a known analytical solution. (Skewed: there exists a non-horizontal and non-vertical symmetry-axis while the load is vertical.)

The paper closes with an example application of the theory to planar tapered beams. This gives a method to obtain a statically possible stress distribution in a trapezoidal plate by transforming the problem along with the loads to a rectangular beam, where the the Euler-Bernoulli beam theory gives a statically possible stress distribution. This solution may be transformed back giving a statically possible solution to the trapezoidal problem.

\section{Preliminaries}\label{sec:2}

The coordinate description of real $n$ dimensional projective space (here $\mathcal{P}^n$) identifies $m$ dimensional projective subspaces of $\mathcal{P}^n$ with $m+1$ dimensional linear subspaces of $\mathbb{R}^{n+1}$. Formally one can introduce the equivalence relation $\vec{x} \sim \vec{y} \iff  \vec{x}=\lambda\vec{y}$ ($\vec{x},\vec{y}\in \mathbb{R}^{n+1}, \lambda\in \mathbb{R}$) and have points of $\mathcal{P}^n$ as equivalence classes of $\mathbb{R}^{n+1} \setminus \{ 0 \}$ with respect to $\sim$. These equivalence classes in themselves capture incidence properties, to be able to handle the metric nature of elasticity we have to fix preferred (appropriately chosen) representants. Let us denote $\mathbb{R}^{n+1}$ with $\mathbb{V}$  for simplicity, chose an orthonormal base $\{\vec{e}_i\}$ in it ($i \in \{0,1,\dots,n \}$) and denote the usual scalar product in it with $\scal{ \ }{ \ }$. Our preferred choice of representants will be such that a finite (not at infinity) point $p_f$ should be represented with vector $\vec{p}_f$ satisfying $\scal{\vec{p}_f}{\vec{e}_0}=1$ while ideal (at infinity) point $p_i$ with vector $\vec{p}_i$ satisfying $\scal{\vec{p}_i}{\vec{e}_0}=0$. We may represent an $m-1$ dimensional projective subspace spanned by points $q_i$ ($i \in \{1,\dots,m\}$) through the exterior product as $ \vec{q}_1 \wedge,\dots,\wedge \vec{q}_m \in \bigwedge^m\mathbb{V}$. The space $\bigwedge^m\mathbb{V}$ is a vector space, it has an orthonormal base determined by $\{\vec{e}_i\}$ that is spanned by $\{\vec{e}_{i_1}\wedge \dots \wedge \vec{e}_{i_m} \ \vert \ 0\leq i_1 <\dots < i_m\leq n \}$. To simplify notation we will omit the wedges and the multiple $\vec{e}$-s, for instance $\vec{e}_{i}\wedge\vec{e}_{j} \wedge \vec{e}_{k}$ will be written as $\vec{e}_{ijk}$ ($i<j<k$). In some cases it will be shorter to write the terms missing, such that $\vec{e}_{\widehat{i_1\dots i_j}}$ will contain all the base vectors in lexicographic order, whose indices are missing from the list $i_i \dots i_j$.

A dual description of projective subspaces is through linear functionals $\vec{\psi}\in \mathbb{V}^\vee$, such that $\vec{\psi}$ represents a projective hyperplane, containing all the points $p$ having representants satisfying $\vec{\psi}(\vec{p})=0$. Lower dimensional subspaces may also be represented through the exterior algebra of $\mathbb{V}^\vee$: given hyperplanes represented by $\vec{\psi}_i$, their intersection is represented by $ \vec{\psi}_1 \wedge,\dots,\wedge \vec{\psi}_m \in \bigwedge^m\mathbb{V}^\vee$, which is an $n-m$ dimensional subspace. The space $\bigwedge^m\mathbb{V}^\vee$ is also a vector space. Denoting the dual base of $\mathbb{V}^\vee$ with $\{\vec{\alpha}_i\}$ such that $\vec{\alpha}_i(\vec{e}_j)=\delta_{ij}$ (the Kronecker-delta), the corresponding orthonormal base in $\bigwedge^m\mathbb{V}^\vee$ is $\{\vec{\alpha}_{i_1 \dots i_m} \ \vert \ 0\leq i_1 <\dots < i_m\leq n \}$.

The Hodge-star operator, mapping $m$-vectors to $(n+1-m)$-vectors can be given on the base vectors as
\begin{align}
\star : \bigwedge_{i\in \mathcal{I}} \vec{e}_i \mapsto (-1)^t \bigwedge_{j\in \mathcal{J}} \vec{e}_j \label{eq:Hodge}
\end{align}
where $\{0, \dots n \}$ is the disjoint union of $\mathcal{I}$ and $\mathcal{J}$ and $t$ is the number of permutations required to bring $\{\mathcal{I},\mathcal{J}\}$ to $(1 \dots n)$. Combining the Hodge-star with the isomorphism $\vec{\alpha}_i \leftrightarrow \vec{e}_i$ gives the map
\begin{align}
\dag : \bigwedge_{i\in \mathcal{I}} \vec{e}_i \mapsto (-1)^t \bigwedge_{j\in \mathcal{J}} \vec{\alpha}_j \label{eq:transition}
\end{align} 
which transitions between the primal and dual representation of the same subspace ($\mathcal{I}$ and $\mathcal{J}$ are as in the case of the Hodge-star). With some possible abuse of notation we will use $\dag$ both in the $\bigwedge^m \mathbb{V} \to \bigwedge^{n+1-m} \mathbb{V}^\vee$ and $\bigwedge^m\mathbb{V}^\vee \to \bigwedge^{n+1-m} \mathbb{V}$ sense. The scalar product of $\vec{x},\vec{y} \in \bigwedge^m\mathbb{V}$ is defined as
\begin{align}
\scal{\vec{x}}{\vec{y}}:=\star(\vec{x} \wedge \star \vec{y})=\star(\star\vec{x} \wedge  \vec{y}).
\end{align}
This scalar product can be used to express the evaluation rule of $\vec{\chi} \in \bigwedge^m \mathbb{V}^{\vee}$ as
\begin{align}
\vec{\chi}(\vec{y})=\scal{\vec{x}}{\vec{y}} \text{ where } \vec{x}=\sum x_{i_1 \dots i_m} \vec{e}_{i_1 \dots i_m} \text{ and } \vec{\chi}=\sum x_{i_1 \dots i_m} \vec{\alpha}_{i_1 \dots i_m}
\end{align} 
holds. \\

The primal description lends itself to naturally represent forces: what would be force $\underline{\vec{F}}$ acting at point $\underline{\vec{x}}$ ($\underline{\vec{x}},\underline{\vec{F}}\in \mathbb{R}^n$), gets expressed as  $\vec{f}=(1,\underline{\vec{x}})\wedge(0,\underline{\vec{F}})$ doubling as projective line coordinates for the line of action of the force. Pure moments can be thought of as having lines of action at infinity. Infinitesimal motions are dual vectors to forces, given small displacement $\vec{\phi} \in \bigwedge^2\mathbb{V}^\vee $ the work of $\vec{f}$ on $\vec{\phi}$ is $\vec{\phi}(\vec{f})$. If $\vec{\phi}$ is a pure rotation it can be written as a product $\vec{\phi}=\vec{\chi}_1 \wedge\vec{\chi}_2$ if it is a pure translation as $\vec{\phi}=\vec{\alpha}_0 \wedge\vec{\chi}$ having the axis (point-wise fix subspace) of rotation lying in the ideal hyperplane $\alpha_0$.\\

Projective transformations of $\mathcal{P}^n$ correspond to equivalence classes of invertible $(n+1) \times (n+1)$ matrices (under equivalence relation $\sim$). In what follows we choose a representant matrix $\vec{T}$ from the appropriate equivalence class and treat it as given. We treat every vector as a column vector and multiply accordingly. We express the geometric constraints through the equivalence relation, denoting the transformed quantities with a prime (as $p \mapsto p'$). 

Point representants $\vec{p}$ and $\vec{p}'$ satisfy
\begin{align}
\vec{p}' \sim \vec{T}\vec{p}.
\end{align}
The effect of $\vec{T}$ on $\bigwedge^m \mathbb{V}$ is given by the $m$-th cofactor matrix \cite{prasolov1994problems} $C_m\vec{T}$ as
\begin{align}
\left(\vec{T} \vec{p}_1 \wedge \dots \wedge \vec{T} \vec{p}_m\right)=C_m\vec{T}\left(\vec{p}_1 \wedge \dots \wedge \vec{p}_m \right)
\end{align}
implying 
\begin{align}
\left(\vec{p}_1 \wedge \dots \wedge \vec{p}_m \right)' \sim C_m\vec{T}\left(\vec{p}_1 \wedge \dots \wedge \vec{p}_m \right).
\end{align}
On functionals $\vec{\chi} \in \bigwedge^m \mathbb{V}^{\vee}$ we will use the transpose of the $m$-th adjugate \cite{aitken2017determinants} $\text{adj}_m\vec{T}^T$  to describe transformations which we will write as
\begin{align}
\vec{\chi}' \sim \text{adj}_m\vec{T}^T \vec{\chi}
\end{align}
for short. This preserves all incidences due to
\begin{align}
\vec{\chi}'(\vec{y}')=\scal{\vec{x}'}{\vec{y}'}=\scal{\text{adj}_m\vec{T}^T\vec{x}}{C_m\vec{T}\vec{y}}=\scal{\vec{x}}{\text{adj}_m\vec{T}C_m\vec{T}\vec{y}}=|\vec{T}|\scal{\vec{x}}{\vec{y}}\label{eq:adj_incidnce}
\end{align}
which can only be $0$ if $\scal{\vec{x}}{\vec{y}}=0$ ($|\vec{T}|$ denotes the determinant of $\vec{T}$). When transforming forces and infinitesimal rotations we have to be stricter than equivalence relations in order to preserve static equilibrium and kinematic compatibility. We have:
\begin{align}
\vec{f}\mapsto & \lambda_f C_2\vec{T} \vec{f} \label{eq:ftrafo} \\
\vec{\phi}\mapsto & \lambda_{\phi} \text{adj}_2\vec{T}^{T} \vec{\phi} \label{eq:phitrafo}
\end{align}
where the scalars $\lambda_f$ and $\lambda_{\phi}$ account for $\vec{T}$ being an arbitrary (fixed) representant. The work of $\vec{f}$ on $\vec{\phi}$ transforms as
\begin{align}
\vec{\phi}(\vec{f})\mapsto \lambda_f \lambda_{\phi} |\vec{T}| \vec{\phi}(\vec{f}).\label{eq:worktrafo}
\end{align}

\section{Algebraic description}
\subsection{Stresses, body forces}
Since we embedded the Euclidean part of the $n$ dimensional projective space in the $\vec{\alpha}_0(\vec{v})=1$ plane of an $n+1$ dimensional vector-space, we will similarly describe $n$ dimensional continua as a thin piece of $n+1$ dimensional material between planes $\vec{\alpha}_0(\vec{v})=1$ and $\vec{\alpha}_0(\vec{v})=1+dt$. The stress state of the material may be described similarly to the usual affine description with an $(n+1) \times (n+1)$ symmetric  Cauchy stress tensor, the matrix representation of which we will denote with $\vec{[\sigma]}$. In the traditional engineering approach this matrix gets multiplied with a normal vector which represents a linear functional through the scalar product and thus for us it is a dual vector. As such we will interpret the stress tensor as a linear map from the dual-space to the vector-space. The fact that we want to describe an $n$-dimensional problem in $n+1$ coordinates means the matrix will satisfy $\vec{[\sigma]}\vec{\alpha}_0=\vec{0}$ and no component of it will depend on $\vec{\alpha}_0$.

\begin{figure}
\begin{center}
\includegraphics[width=0.4\textwidth]{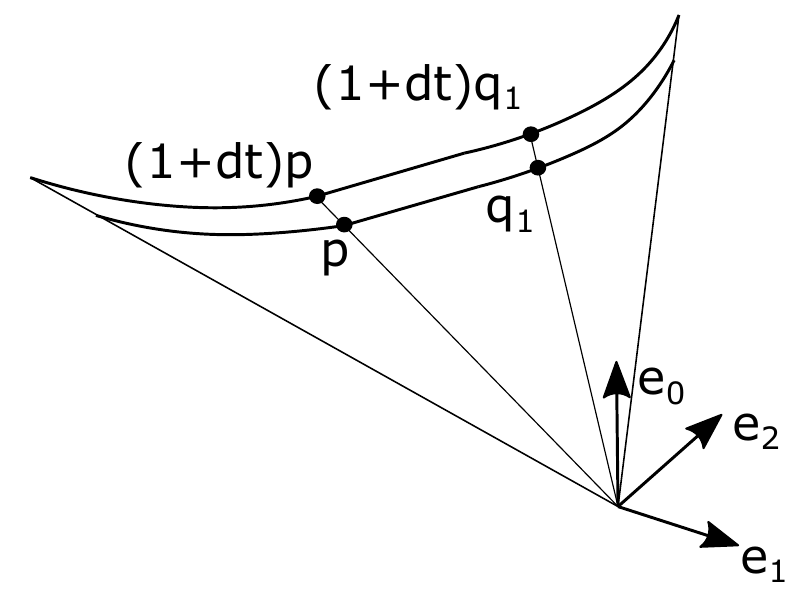}
\end{center}
\caption{For the projective planar case we consider a $3$ dimensional solid with thickness $dt<<1$. What appears as a cut along a line segment in the projective problem is a trapezoid cut surface in the higher dimensional affine problem.}\label{fig:d}
\end{figure}

We may describe stresses at point $p$ by cutting the material in half, considering a small patch of the cut surface at $p$ and the resultant of the force system acting on that patch. What appears in the projective setting as an $(n-1)$ simplicial patch with vertices $p$ and $q_i$ ($i \in \{1,2,\dots,n-1\}$) is in the affine setting an $n$ dimensional polytope, with vertices $\vec{p},\vec{q}_i,(1+dt)\vec{p},(1+dt)\vec{q}_i$ (the projective planar case is drawn in Figure \ref{fig:d}). As in the projective problem $q_i \to p$ such that $dt$ and all angles of the (simplex) at $p$ stay constant, the $n$ dimensional polytope tends to a prism having the simplex as a base. The $n-1$ volumes of the simplex-based faces of the polytope are given by the coordinates of   
\begin{align}
\frac{1}{(n-1)!}\left(\vec{q}_1-\vec{p}\wedge,\dots,\wedge\vec{q}_{n-1}-\vec{p}\right)
\end{align}
and we can also observe 
\begin{align}
\vec{\alpha}_0(\vec{q}_i-\vec{p})=0 \implies  \vec{\alpha}_0 \wedge \vec{\alpha}_{i_2} \wedge \dots \wedge \vec{\alpha}_{i_{n-1}} \left(\vec{q}_1-\vec{p}\wedge,\dots,\wedge\vec{q}_{n-1}-\vec{p} \right)=0
\end{align}
holds for all functionals starting with $\vec{\alpha_0}$. As such 
\begin{align}
\frac{1}{(n-1)!} dt\vec{p}\wedge\left(\vec{q}_1-\vec{p}\wedge,\dots,\wedge\vec{q}_{n-1}-\vec{p}\right)= \frac{1}{(n-1)!}dt\vec{p}\wedge\vec{q}_1\wedge,\dots,\wedge\vec{q}_{n-1}
\end{align}
will contain the first $n$ coordinates the $n$-volumes of the faces of the prism that stand on the faces of the simplex. As such introducing the dual vector
\begin{align}
\vec{\nu}:=\frac{1}{(n-1)!}\dag\left(dt\vec{p}\wedge\vec{q}_1\wedge,\dots,\wedge\vec{q}_{n-1}\right) \label{eq:nudef}
\end{align}
and denoting the stress tensor at point $p$ with $\vec{[\sigma_p]}$ as $q_i \to p$ one can approximate the contact force on the patch to arbitrary precision with $d\vec{F}\approx\vec{[\sigma_p]}\vec{\nu}$. Although $d\vec{F}$ acts at the centroid at the contact patch and not in $p$, when writing up the moment with respect to the origin the effect of the difference in point of action will be quadratic in the sides of the parallelepiped and thus negligible compared to the term $\vec{p}\wedge d \vec{F}$.

As such for the moment of the stresses we keep only the term $\vec{p}\wedge d \vec{F}$ which we will calculate through matrix-multiplication. There is a unique matrix $\vec{[p\wedge]}$, such that $\vec{p} \wedge \vec{x}=\vec{[p\wedge]}\vec{x}$ for all $\vec{x} \in \bigwedge^m\mathbb{V}$ (for a fixed $m$). For us it will suffice to look at the $m=1$ case. Labelling the base-vectors of the image space of $2$-vectors with index pairs $ij $ (such that $i<j$ holds), coordinate-wise the entry in the $ij$-th row and $k$-th column of the matrix can be given as
\begin{align}
[p\wedge]_{ij,k}=\begin{cases}
p_i \quad &\text{if } k=j\\ 
-p_j \quad &\text{if } k=i\\
0 \quad &\text{otherwise.}
\end{cases}
\end{align}

We will also use the transpose of this matrix acting on dual coordinates as $\vec{[p\wedge]}^T: \ \bigwedge^{2}\mathbb{V}^\vee \to \bigwedge\mathbb{V}^\vee $. With it we have  
\begin{lemma}\label{lem:1}
$\dag\left(\vec{p}\wedge\vec{q}_1\wedge,\dots,\wedge\vec{q}_{n-1}\right)=(-1)^{n+1}\vec{[p\wedge]}^T\dag\left(\vec{q}_1\wedge,\dots,\wedge\vec{q}_{n-1}\right)$
\end{lemma}
\begin{proof}
Let us introduce notation $\vec{r}=\vec{q}_1\wedge,\dots,\wedge\vec{q}_{n-1}$. As $\vec{r}$ is an $n-1$-vector we can more efficiently label the coordinates with the base vectors missing instead of the base vectors present as:
\begin{align}
\vec{r}=\sum_{i<j} r_{\widehat{ij}} \vec{e}_{\widehat{ij}}.
\end{align}

With this we have on one hand
\begin{align}
\dag\left(\vec{p}\wedge \vec{r}\right) &=\dag\left(\sum_{i=0}^{n-1}(-1)^i p_i r_{\widehat{ij}} \vec{e}_{\hat{j}} + \sum_{j=1}^{n} (-1)^{j-1}p_j r_{\widehat{ij}} \vec{e}_{\hat{i}} \right)\\
&=\left( \sum_{i=0}^{n-1} (-1)^{i+j+n} p_i r_{\widehat{ij}} \vec{\alpha}_{j} + \sum_{j=1}^{n} (-1)^{i+j+1+n} p_j r_{\widehat{ij}} \vec{\alpha}_{i}  \right)\label{eq:lem1_1} 
\end{align}
and on the other hand
\begin{align}
\dag \vec{r}&=\sum_{i<j} (-1)^{i+j-1} r_{\widehat{ij}} \vec{\alpha}_{ij}\\
(-1)^{n+1} \vec{[p\wedge]}^T  \dag \vec{r} &= \sum_{i=0}^{n-1} (-1)^{i+j+n} p_i r_{\widehat{ij}} \vec{\alpha}_{j} + \sum_{j=1}^{n} (-1)^{i+j+n+1} p_j r_{\widehat{ij}} \vec{\alpha}_{i}. \label{eq:lem1_2}
\end{align}
Comparing Equations \eqref{eq:lem1_1} and \eqref{eq:lem1_2} completes the proof.
\end{proof}

We define the projective stress tensor (in matrix form, corresponding to point $p$) as
\begin{align}
\vec{[\Sigma_p]}:=dt(-1)^{n+1}\vec{[p\wedge]} \vec{[\sigma_p]}\vec{[p\wedge]}^T
\end{align}
where $dt=1$ may be used for practical purposes. The stress matrix has a structure and can be thought of as a block matrix, namely
\begin{align}
\vec{[\Sigma_p]}=
\begin{bmatrix}
\vec{[\overline{\sigma_p}]} & \vec{[\mu]}^T \\
\vec{[\mu]} &  \vec{[\zeta]}
\end{bmatrix}.
\end{align}
To see this, recall that $\vec{[\Sigma_p]}=\vec{[\Sigma_p]}^T$ and it represents a map acting on dual line coordinates returning primal line coordinates. Thus for any pair of base vectors of the form $\vec{e}_0\wedge\vec{e}_i$ and $\vec{e}_0\wedge\vec{e}_j$ the expression 
$\scal{\vec{e}_0\wedge\vec{e}_i}{\vec{[\Sigma_p]} \dag (\vec{e}_0\wedge\vec{e}_j})$ is the $\vec{e_i}$ directional component of $d\vec{F}$ if the cut is orthogonal to $\vec{e}_j$ and has unit area. Thus $\vec{[\overline{\sigma_p}]}$ is the restriction of $\vec{\sigma_p}$ to $\text{span} (\vec{\alpha}_1, \dots \vec{\alpha}_n )$. The block $\vec{[\mu]}$ represents the moments of the stresses and its rows are linear combinations of the rows of $\vec{[\overline{\sigma_p}]}$ according to how moments are defined. Due to the incidence condition $\vec{p} \wedge d\vec{f}=\vec{0}$ if $\vec{[\sigma_p]}$ is non-degenerate the rank of $\vec{[\Sigma_p]}$ has to be $n$; thus the rows of $\vec{[\zeta]}$ are linear combinations of the rows of $\vec{[\mu]}^T$ with the same coefficients as in the case of $\vec{[\overline{\sigma_p}]}$ and $\vec{[\mu]}$.\\

As body forces $\underline{\vec{g}} \in \mathbb{R}^n $ are functions that give a force when integrated on an $n$ dimensional domain, we will treat them similarly to forces and have them of shape $\vec{g}=(0,\underline{\vec{g}})$.\\

When considering projective transformations of stresses and body forces we have two requirements. One is to preserve the geometric structure of the problem including lines of action of forces, and the other is to preserve static equilibrium. Preserving the geometry is done by finding the correct equivalence classes of matrices and vectors, while preserving static equilibrium is done by choosing the correct representatnt from these equivalence classes. To prescribe transformation rules fulfilling these two requirements we will consider a discretization the refinement of which converges to the continuum description.

Given a continuum problem having domain  $\mathcal{D} \subset \mathcal{P}^n$ it is in static equilibrium exactly if for any sub-domain $\mathcal{S} \subset \mathcal{D}$ the body forces in $\mathcal{S}$ together with the stresses on the boundary of $\mathcal{S}$ are in static equilibrium. Let us choose $N$ points randomly, from a uniform distribution above $\mathcal{S}$. Let us also create a simplicial cover of $\mathcal{S}$ by triangulating these points. Let us label the $n$-simplices with indices $l\in \mathcal{L}$ and the $(n-1)$-simplices on the boundary of $\mathcal{S}$ with indices $m \in \mathcal{M}$. On each boundary $(n-1)$-simplex we will have a force resultant $d\vec{f}_m$ and to each $n$-simplex corresponds a body-force $d\vec{b}_l$ (both are line-coordinates). We know from line geometry that static equilibrium will be preserved through
\begin{align}
\sum_{l\in \mathcal{L}} \lambda_f C_2\vec{T} d\vec{b}_l+\sum_{m\in \mathcal{M}} \lambda_f C_2\vec{T} d\vec{f}_m= \lambda_f C_2\vec{T} \left( \sum_{l\in \mathcal{L}} d\vec{b}_l +\sum_{m\in \mathcal{M}} d\vec{f}_m \right)=\lambda_f C_2\vec{T} \vec{0}=\vec{0}.
\end{align}
Figuring out the transformation rule for the stress tensor and body-force vector requires that if one simplicial approximation converges to a continuum solution than the transformed approximation has to converge to the transformed continuum solution.  

To describe the transformation of the stress tensor consider an $(n-1)$-simplex on the boundary of $\mathcal{S}$ and assign the roles of points $p$ and $q_i$ in Equation \eqref{eq:nudef} to the vertices of the simplex and construct $\vec{r}_m=\bigwedge_{i=1}^{n-1} \vec{q}_i$. We may approximate the force-resultant on the simplex as $d\vec{f}_m \approx \frac{1}{(n-1)!} \vec{[\Sigma_{p_m}]}\dag\vec{r}_m$ the error of which will be arbitrarily small as $N\to \infty$. Similarly, the transformed stress tensor will be the one where the approximation
\begin{align}
\frac{1}{(n-1)!}\vec{[\Sigma_p]}' \left( \dag \vec{r} \right)' \approx \lambda_f C_2\vec{T} d\vec{f}_m \label{eq:show_1}
\end{align}
gets arbitrarily precise as $N\to \infty$. When expressing $\left( \dag \vec{r} \right)'$ we have to keep in mind that $\dag \vec{r}$ was built from representants satifying $\vec{\alpha_0}(\vec{q}_i)=1$. Introducing $\lambda_{q_i}=\vec{\alpha}_0(\vec{T}\vec{q}_i)$ we may write
\begin{align}
\left( \dag \vec{r} \right)' &= \dag \left( \lambda_{q_1}^{-1} \vec{T} \vec{q}_1 \wedge,\dots,\wedge \lambda_{q_{n-1}}^{-1} \vec{T} \vec{q}_{n-1}\right)\\
&= \left( \Pi_{i=1}^{n-1} \lambda_{q_i}^{-1} \right) \dag \left(  \vec{T} \vec{q}_1 \wedge,\dots,\wedge \vec{T} \vec{q}_{n-1}\right)\\
&=  \left( \Pi_{i=1}^{n-1} \lambda_{q_i}^{-1} \right)\dag \left( C_{n-1}\vec{T} \left( \vec{q}_1 \wedge,\dots,\wedge \vec{q}_{n-1}\right) \right)\\
&=\left( \Pi_{i=1}^{n-1} \lambda_{q_i}^{-1} \right) \text{adj}_{2}\vec{T}^T \dag \vec{r}. 
\end{align}
As $N \to \infty$ we will have $\lambda_{q_i} \approx \lambda_p$ from which
\begin{align}
\vec{[\Sigma_p]}'=\frac{\lambda_f \lambda_p^{(n-1)}}{|\vec{T}|}C_2\vec{T}\vec{[\Sigma_p]}C_2\vec{T}^T \label{eq:fesztrafo}
\end{align}
follows, where $\lambda_p=\vec{\alpha}_0(\vec{T}\vec{p})$ is the scalar one has to scale $\vec{T}\vec{p}$ back with in order to get a preferred representant.\\

To describe the transformation of the body-force vector consider an $n$-simplex, and label its vertices with $\{p,q_1,\dots,q_n\}$. If $\vec{g}_p$ is the body-force in $p$ and the $n$-volume of the simplex is $V_{n,l}$ we may approximate the body-force resultant in the simplex as $d\vec{g}_l \approx V_{n,l} \vec{g}_{p}$ and express the effect of the body-force corresponding to the simplex as $d\vec{b}_l \approx \vec{p} \wedge d\vec{g}_l$. Again as $N\to \infty$ this approximation gets arbitrarily precise
 and the transformed body-force vector will be the one where
\begin{align}
\lambda_f C_2\vec{T} d\vec{b}_l \approx \vec{p}' \wedge  V'_{n,l} d \vec{g}'_p \label{eq:show_3}
\end{align}
also holds for arbitrary precision.

We look for a transformed body force of shape $\vec{g}'_p=(0,\underline{\vec{g}'_p})$, that is we require $\vec{\alpha}_0(\vec{g}'_p)=0$. The transformation rule of line coordinates implies
\begin{align}
C_2\vec{T}(\vec{p} \wedge d\vec{g}_l)=\vec{T}\vec{p} \wedge \vec{T} d\vec{g}_l= \vec{p}' \wedge \lambda_p \vec{T} d\vec{g}_l
\end{align} 
where in general $\vec{\alpha}_0(\vec{T} d\vec{g}_l)\neq 0$. Relying on the antisymmetry of the wedge product we set 
\begin{align}
d\vec{g}'_l=\lambda_p \vec{T}d\vec{g}_l-\vec{\alpha}_0(\vec{T}d\vec{g}_l)\vec{T}\vec{p} \approx  V_{n,l} \left( \lambda_p \vec{T}\vec{g}_p-\vec{\alpha}_0(\vec{T}\vec{g}_p)\vec{T}\vec{p} \right)  \label{eq:sinto}
\end{align}
satisfying $C_2\vec{T}(\vec{p} \wedge d\vec{g}_l)=\vec{p}' \wedge d\vec{g}'_l$ and $\vec{\alpha}_0(d\vec{g}'_l)=0$.
We will now use the fact that $\alpha_0(\vec{p})=1=\alpha_0(\vec{q}_k ) \quad \forall k\in \{1 \dots n\}$ and and express the $n$-volumes of the simplices as $V_{n,l}=V_{n+1,l}/(n+1)!$ where $V_{n+1,l}$ is the volume of the $(n+1)$-simplex formed by adding the origin of $\mathbb{R}^{n+1}$ as a vertex to the set of vectors $\{\vec{p},\vec{q}_1,\dots,\vec{q}_n \}$. Similarly we will have $V'_{n,l}=V'_{n+1,l}/(n+1)!$ for the volumes of the transformed simplices. As the transformation $\vec{p} \mapsto \vec{p}'$ consists of a linear map and scaling the volumes will transform as 
\begin{align}
V'_{n+1,l}=\frac{|\vec{T|}\lambda_p^{-1}}{\Pi_{k=1}^{n} \vec{\alpha}_0(\vec{T}\vec{q}_{k})} V_{n+1,l}=\frac{|\vec{T|}\lambda_p^{-1}}{\Pi_{k=1}^{n} \lambda_{q_k}} V_{n+1,l}.
\end{align}

By substituting this into Equation \eqref{eq:sinto} and simplifying we arrive at 
\begin{align}
\frac{|\vec{T|}\lambda_p^{-1}}{\Pi_{k=1}^{n} \lambda_{q_k}} \vec{g}'_{p}\approx \lambda_p \vec{T}\vec{g}_{p}-\vec{\alpha}_0(\vec{T}\vec{g}_{p})\vec{T}\vec{p}.
\end{align} 
As $N \to \infty$ we will have $\lambda_{q_k} \approx \lambda_p \quad \forall k\in \{1 \dots n\}$ implying that the transformation rule of the body force at $p$ is
\begin{align}
\vec{g}'_p=\frac{\lambda_p^{n+1}}{|\vec{T|}} \left(  \lambda_p \vec{T}\vec{g}_{p}-\vec{\alpha}_0(\vec{T}\vec{g}_{p})\vec{T}\vec{p} \right)
\end{align}
with which the discretization will satisfy Equation \eqref{eq:show_3} to arbitrary precision, as $N \to \infty$.

\subsection{Strains}
Strains of the material correspond to the relative translations of the two endpoints $p$ and $q$ of an infinitesimally short chord of the material. Traditionally translations correspond to rotations around $n-2$ dimensional subspaces at infinity (lying in the ideal hyperplane), but multiple infinitesimal rotations can cause the same displacement of $q$ relative to $p$. We will have to use this property, otherwise we would only be able to describe affine transformations leaving the ideal hyperplane fixed. In general the strain tensor corresponding to point $p$ will give rotations lying in a specific hyperplane $\psi_{ref}$ that does not contain point $p$. We will denote this hyperplane with a superscript and later we will give a method to change the reference-hyperplane. We will start by introducing the strain tensor in the traditional setting where $\psi_{ref}=\alpha_0$, that is the reference plane is the ideal hyperplane.\\

To introduce the projective strain tensor (matrix) we again treat the problem as an $n+1$ dimensional one. We have the affine strain tensor corresponding to point $p$ in matrix form as $\vec{[\epsilon_p  ^{\alpha_0}]}$ satisfying $\vec{[\epsilon_p^{\alpha_0}]}\vec{e}_0=0$ and no component of it depending on $\vec{e}_0$. If the two endpoints of an infinitesimally short chord of the material are $p$ and $q$ (with properly chosen representants $\vec{p}$ and $\vec{q}$), we have $(0,\vec{\underline{\Delta}})=\vec{[\epsilon_p ^{\alpha_0}]}(\vec{q}-\vec{p})$. If we chose the reference-hyperplane to be the hyperplane at infinity, the translation may be expressed as $d\vec{\phi}=\vec{\alpha}_0 \wedge \vec{[\epsilon_p ^{\alpha_0}]}(\vec{q}-\vec{p})$. Motivated by this we define the projective strain tensor as
\begin{align}
\vec{[E_p^{\alpha_0}]}:=\vec{[\alpha_0\wedge]} \vec{[\epsilon_p ^{\alpha_0}]}\vec{[\alpha_0\wedge]}^T
\end{align}
that will act on primal line coordinates in the form of $\vec{p}\wedge \vec{q}$. To justify this we may observe that the only non-zero elements of $\vec{[\alpha_0\wedge]}^T$ are $[\alpha_0\wedge]^T_{j,0j}=1$ where $j\in\{1,\dots,n\}$. Thus we have $(\vec{q}-\vec{p})=\vec{[\alpha_0\wedge]}^T (\vec{p}\wedge \vec{q})$.

If the reference plane is the ideal hyperplane the strain matrix has the block structure
\begin{align}
\vec{[E_p^{\alpha_0}]}=
\begin{bmatrix}
\vec{[\overline{\epsilon_p}]} & \vec{\left[0_{n \times \binom{n}{2}}\right]} \\
\vec{\left[0_{\binom{n}{2} \times n}\right]} &  \vec{\left[0_{\binom{n}{2} \times \binom{n}{2}}\right]}
\end{bmatrix}
\end{align} 
which can be seen through a similar reasoning as in the case of $\vec{[\Sigma_p]}$. To describe the change of reference-hyperplane we will need the following observation:

\begin{lemma}\label{lem:2}
If $\psi_{ref}$ is not the ideal hyperplane there is a linear map $L: \ \mathbb{V}^\vee \rightarrow \mathbb{V}^\vee$ describing the change of reference-hyperplane, corresponding to a central-axial collineation of $\mathcal{P}^n$ with center $p$.
\end{lemma}
  
\begin{proof}
Since $p$ does not lie on $\psi_{ref}$ we may chose a representant satisfying $\vec{\psi}_{ref}(\vec{p})=1$. One can choose a generator-system $\{ \vec{\alpha}_0, \vec{\psi}_i \}$ with $i\in \{1,\dots,n\}$ such that $\vec{\psi}_i(\vec{p})=0$ holds. The map is determined by  equations
\begin{align}
\vec{L}\vec{\alpha}_0 & =   \vec{\psi}_{ref} \\
\vec{L}\vec{\psi}_i  & = \vec{\psi}_i
\end{align}
and $\vec{L}$ represents a central axial collineation since all hyperplanes incident with point $p$ are fixed. Now consider point $p$ (with representant  $\vec{p}=(1,\underline{\vec{p}})$) undergoing translation $\vec{\underline{\Delta}}\in \mathbb{R}^n$. The translation describing this motion can be given as $\vec{\alpha}_0\wedge \vec{\Delta} $ where $\vec{\Delta}=(-\scal{\underline{\vec{\Delta}}}{\underline{\vec{p}}},\vec{\underline{\Delta}})$ represents a hyperplane incident with $p$. This can be seen by choosing unit vector $\underline{\vec{u}}\in \mathbb{R}^n$ and the corresponding ideal point $\vec{u}=(0,\underline{\vec{u}})$ and calculating the $\vec{u}$ directional component of the translation as 
\begin{align}
(\vec{\alpha}_0\wedge \vec{\Delta})(\vec{p}\wedge \vec{u})=
\begin{vmatrix}
\vec{\alpha}_0(\vec{p}) & \vec{\alpha}_0(\vec{u}) \\
\vec{\Delta}(\vec{p}) & \vec{\Delta}(\vec{u})
\end{vmatrix}
=\vec{\alpha}_0(\vec{p}) \vec{\Delta}(\vec{u})=\scal{\underline{\vec{\Delta}}}{\underline{\vec{u}}}
\end{align}
(recall that $\vec{\Delta}(\vec{p})=0$).
The image of this motion is the rotation $\vec{\psi}_{ref} \wedge \vec{\Delta} $ which causes the same translational motion at $p$ as 
\begin{align}
(\vec{\psi}_{ref}\wedge \vec{\Delta})(\vec{p}\wedge \vec{u})=
\begin{vmatrix}
\vec{\psi}_{ref}(\vec{p}) & \vec{\psi}_{ref}(\vec{u}) \\
\vec{\Delta}(\vec{p}) & \vec{\Delta}(\vec{u})
\end{vmatrix}
=\vec{\psi}_{ref}(\vec{p}) \vec{\Delta}(\vec{u})=\scal{\underline{\vec{\Delta}}}{\underline{\vec{u}}}
\end{align}
since we had $\vec{\psi}_{ref}(\vec{p})=1$.
\end{proof}

According to the representation of transformations laid out in Section \ref{sec:2}, as $\vec{L}$ acts on hyperplane coordinates the the map on point coordinates is given by $\vec{A}=\sqrt[n]{|\vec{L}|} \vec{L}^{-T}$ satisfying $\vec{L}=\text{adj} \vec{A}$. Here we specifically deviate and neglect the $\sqrt[n]{|\vec{L}|}$ term, since this neglection preserves the geometric constrains and will keep the expression of the work reference-hyperplane independent. Thus we will write $\vec{[\epsilon_p ^{\vec{L}\vec{\alpha}_0}]} =  \vec{L}^{-1} \vec{[\epsilon_p ^{\alpha_0}]}\vec{L}^{-T}$. Finally, changing the incidence conditions to match the reference-hyperplane we have
\begin{align}
\vec{[E_p^{\vec{L}\alpha_0}]}=\vec{[\vec{L}\alpha_0\wedge]} \vec{[\epsilon_p ^{\vec{L}\alpha_0}]}\vec{[\vec{L}\alpha_0\wedge]}^T
\end{align}
which satisfies the 
\begin{align}
\vec{[E_p^{\vec{L}\alpha_0}]}=C_2 \vec{L} \vec{[E_p^{\alpha_0}]} C_2 \vec{L}^T
\end{align}
transformation rule under change of reference-hyperplane. With this in hand we may now derive the transformation rule of the strain tensor under projective transformations. Consider a closed curve in the material, and discretize it to line segments by uniformly distributing points $p_i$ $i\in \{0,\dots,N\}$  on it, according to arclength. As the curve is closed, we have $p_0=p_N$. At each point $p_i$ we may approximate the relative displacement of $p_i$ and $p_{i+1}$ with $\vec{[E_p]}(\vec{p}_i\wedge\vec{p}_{i+1})$. As $N \to \infty$ the approximation gets arbitrarily precise. The compatibility is expressed as the sum of the displacements being the zero vector, which is preserved under the projective transformation as
\begin{align}
\sum_{i=0}^N \lambda_{\phi} \text{adj}_2^T\vec{[E_{p_i}]}(\vec{p}_i\wedge\vec{p}_{i+1})= \lambda_{\phi} \text{adj}_2^T \sum_{i=0}^N \vec{[E_{p_i}]}(\vec{p}_i\wedge\vec{p}_{i+1})=\vec{0}. 
\end{align}
The strain tensor should transform such that the transformed small displacement may be approximated with the transformed strain tensor and the transformed line segment, that is \begin{align}
\lambda_{\phi} \text{adj}_2^T\vec{[E_{p_i}]}(\vec{p}_i\wedge\vec{p}_{i+1}) \approx \vec{[E_{p_i}]'}(\vec{p}'_i\wedge\vec{p}'_{i+1})  \label{eq:show_2}
\end{align}
holds for arbitrary precision as $N \to \infty$. We have \begin{align}
\vec{p}'_i\wedge\vec{p}'_{i+1}=\lambda_{p_i}^{-1} \lambda_{p_{i+1}}^{-1} C_2\vec{T}(\vec{p}_i\wedge\vec{p}_{i+1}) \approx \lambda_{p_i}^{-2} C_2\vec{T}(\vec{p}_i\wedge\vec{p}_{i+1})
\end{align}
since $ \lambda_{p_{i+1}}^{-1} \xrightarrow{N \to \infty} \lambda_{p_i}^{-1}$, which gives 
\begin{align}
\vec{[E_p]}'=\frac{\lambda_{\phi} \lambda_p^{2}}{|\vec{T}|}\text{adj}_2\vec{T}^T\vec{[E_p]}\text{adj}_2\vec{T} \label{eq:epstrafo}
\end{align} through some manipulation.

\subsection{Work}
The work of stresses on strains may be expressed from the matrices above through the Hilbert-Schmidt scalar product as
\begin{align}
W=\text{trace}\left(\vec{[E_p]}^T \vec{[\Sigma_p]}\right) \label{eq:workdef}
\end{align}
which does not depend on the choice of $\psi_{ref}$. To see the independence we recall that $\vec{L}$ was created not to alter the effect of infinitesimal rotation $d\vec{\phi}$ on lines of the shape $\vec{p}\wedge \vec{q}$. That is for any $d\vec{\phi}$:
\begin{align}
\scal{d\vec{\phi}}{\vec{p}\wedge \vec{q}}=\scal{C_2\vec{T}d\vec{\phi}}{\vec{p}\wedge \vec{q}}=\scal{d\vec{\phi}}{C_2 \vec{L}^T(\vec{p}\wedge \vec{q})}
\end{align}  
holds, implying $C_2 \vec{L}^T(\vec{p}\wedge \vec{q})=\vec{p}\wedge \vec{q}$. As the columns of $\vec{[\Sigma_p]}$ are exactly of shape $\vec{p}\wedge \vec{q}$ we have $C_2 \vec{L}^T \vec{[\Sigma_p]} =\vec{[\Sigma_p]}$. Similarly, as each entry of $\vec{[E_p]}^T \vec{[\Sigma_p]}$ is a scalar product of shape $\scal{d\vec{\phi}}{\vec{p}\wedge \vec{q}}$ we have 
$C_2\vec{L}\vec{[E_p^{\vec{\alpha_0}}]}^T \vec{[\Sigma_p]}=\vec{[E_p^{\vec{\alpha_0}}]}^T \vec{[\Sigma_p]}$.
Thus $C_2\vec{L}\vec{[E_p^{\vec{\alpha_0}}]}^T C_2\vec{L}^T \vec{[\Sigma_p]}=\vec{[E_p^{\vec{\alpha_0}}]}^T  \vec{[\Sigma_p]}$ and the expression of the work is indeed reference-hyperplane independent. The correctness of Equation \eqref{eq:workdef} may be seen from the block structure of the matrices, considering the form where the reference-hyperplane is the hyperplane at infinity.

Under projective transformations the work of the transformed stresses on the transformed strains may be expressed from the original as
\begin{align}
W'=&\frac{\lambda_f \lambda_p^{(n-1)}}{|T|} \frac{\lambda_{\phi} \lambda_p^{2}}{|T|}\text{trace}\left(\left(C_2\vec{T}\vec{[\Sigma_p]}C_2\vec{T}^T \right)^T \text{adj}_2\vec{T}^T\vec{[E_p]}\text{adj}_2\vec{T}\right)\\
W'=&\frac{\lambda_f \lambda_{\phi}\lambda_p^{(n+1)}}{|T|^2}|T|^2 \text{trace}\left(C_2\vec{T}\vec{[\Sigma_p]}^T\vec{[E_p]}C_2\vec{T}^{-1}\right)\\
W'=&\lambda_f \lambda_{\phi}\lambda_p^{(n+1)}W.
\end{align}

In the special case of affine transformations it is possible to chose $\vec{T}$ such that $\lambda_p=1$ for all $p$ and if $\lambda_{\phi}\lambda_{f}=1$ then the (point-wise) work is invariant, returning the result of Ostrosablin \cite{affinetransformations_2006_Ostrosablin}. In this case the determinant in equation \eqref{eq:worktrafo} will correspond to the transformation induced change in the volume of the solid. 

\section{Geometric interpretation}
\subsection{Invariant mechanical properties}
As projective transformations are defined to preserve incidences we will investigate what mechanical properties we can relate to incidence. The stress tensor maps vectors representing $n-2$ dimensional projective subspaces to vectors representing lines; while the strain tensor maps vectors representing lines to vectors representing $n-2$ dimensional projective subspaces. If the projective subspaces represented by the starting and the image vectors intersect, the corresponding mechanical phenomenon is that there is no normal stress on the cut determined by the $n-1$ dimensional subspace and there is no normal strain in the direction determined by the line. As such we have

\begin{corollary}\label{thm:cor1}
At point $p$ the absence of normal stresses / strains  along a hyperplane / line containing $p$ is a projective invariant.
\end{corollary}

We may also note, that the kernels of the tensors correspond to projective subspaces and thus we have the second corollary

\begin{corollary}\label{thm:cor2}
At point $p$ the absence of stresses / strains  along a hyperplane / line containing $p$ is a projective invariant.
\end{corollary}

\begin{figure}
\centering
\includegraphics[width=0.4\textwidth]{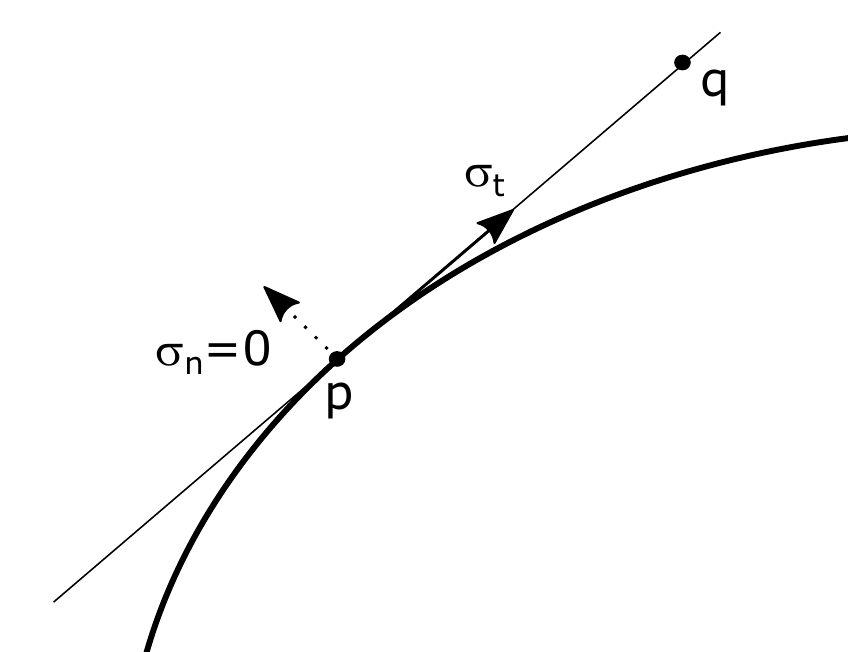}
\includegraphics[width=0.4\textwidth]{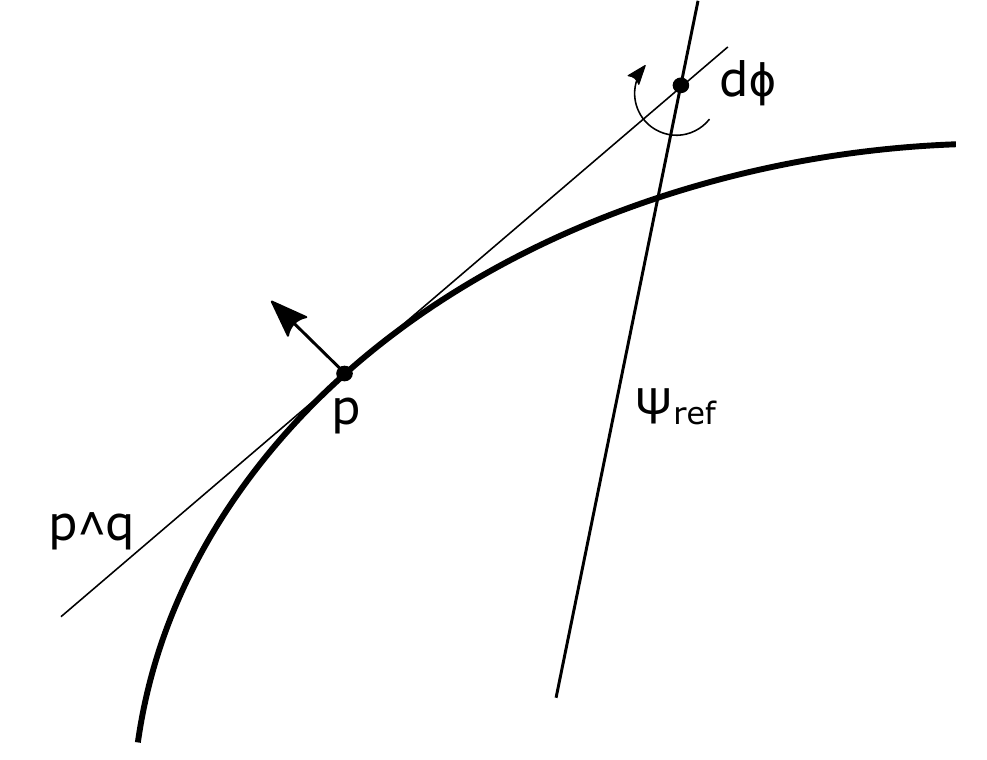}
\caption{Left: if the stress state corresponds to a real conic, it consists of lines tangent to cut directions where no normal stresses occur. Right: if the strain state corresponds a dual conic having real points they line on the tangents where no normal strain occurs. The deformation of point $p$ relative to its neighbouring points on the curve appears normal to the curve.}\label{fig:bc}
\end{figure}

The algebraic description of these corollaries may be done through the symmetric bilinear form determined by the matrices of the tensors. As this is a natural description of quadrics, we may naturally associate a quadric to each stress and strain tensor. In the general case these quadrics live in $\mathcal{P}^{\binom{n+1}{2}-1}$ but for $n=2$ this space may be identified with the plane of the problem, while in $n=3$ the corresponding $\mathcal{P}^5$ is inhabited by the Klein quadric and the geometry of lines of $\mathcal{P}^3$ has been investigated through it. As such we will look at these two cases in detail. An illustration of the two dimensional cases is drawn in Figure \ref{fig:bc}.

\subsection{Generalized conic sections in the plane}
For the stress state in the plane the we may describe the generalized conic section as the set of points $\vec{q}$ satisfying $\scal{\dag\vec{q}}  {\vec{[\Sigma_p]} \dag\vec{q}}=0$. Depending on the eigenvalues of $\vec{[\overline{\sigma_p}]}$ we will see different types of conic sections. Excluding the case where both of the eigenvalues are $0$ since in this case the solution of the bilinear form would be the entire plane, the possible cases are:
\begin{itemize}
\item The conic is a single point. This happens if the principal stresses are non-zero and have the same sign since in this case there are always normal stresses. The point is $p$ as $\vec{[\Sigma_p]} \dag \vec{p}=\vec{0}$.
\item The conic is a pair of intersecting lines. This happens if the principal stresses are non-zero and differ in sign, allowing for directions with no normal stresses. The lines intersect at point $p$. 
\item The conic is a single line. This happens if exactly one of the principal stresses is zero, there is only one direction in which no normal stress occurs.
\end{itemize}

For the planar strain state the corresponding tensor is a linear map acting on line coordinates, thus we give the strain-conic as a dual conic \cite{richter2011perspectives}. In case of non-degenerate dual conics these lines would be tangents of a regular conic. Unfortunately the degenerate case is a bit less straightforward, but the possibilities are also known. Based on the literature on dual conics the relevant possibilities of the dual conic are:
\begin{itemize}
\item A real double line. This happens if the principal strains are non-zero and have the same sign. The line is the reference line $\psi_{ref}$. 
\item Two real points and a real double line on them. This happens if the principal strains are non-zero and differ in sign. The line is again $\psi_{ref}$, the two points correspond to the two directions where the normal strain is $0$. 
\item A real double line and a real double point through it. This happens if exactly one of the principal strains is zero. The line is again the one represented by $\psi_{ref}$, the point is corresponding to the direction in which the normal strain is zero.
\end{itemize}

\subsection{Generalized quadrics in $\mathcal{P}^5$}
For 3 dimensional problems both the stress and strain tensors map line coordinates to line coordinates which may be thought of as points (the Plücker-embedding) or hyperplanes (the dual of the Plücker-embedding) in $\mathcal{P}^5$. We see no benefit of using the dual embedding for any of the two, we are interested in the geometry of the lines. We recall \cite{csikos-kiss} that under the Plücker-embedding lines of $\mathcal{P}^3$ are mapped to points of a hyperbolic quadric in $\mathcal{P}^5$, that is known as the Klein quadric. Given $p\in \mathcal{P}^3$ lines incident with it are mapped to 2 a dimensional projective subspace of $\mathcal{P}^5$ called a Latin plane. Given $\psi_{ref} \subset \mathcal{P}^3$ lines lying in it are mapped to a 2 a dimensional projective subspace of $\mathcal{P}^5$ called a Greek plane. These planes are parts of the kernels of the stress and strain quadrics, which we will give as cones having planar conics as bases and planes as apexes.

In case of the stress tensor the line of action of $d\vec{f}$ has to pass through $p$, we have $ \forall \vec{q}: \ \vec{[\Sigma_p]} \dag \left(\vec{p} \wedge \vec{q} \right) =\vec{0}$. Thus the stress quadric contains as apex the Latin plane corresponding to the lines passing through $p$. We chose the Greek plane corresponding to all the ideal lines to contain the base of the cone, as each ideal line determines a cut going through $p\in \mathcal{P}^3$. This way the problem reduces to finding the zeros of $\scal{\vec{n}}{\vec{[\overline{\sigma_p}]}\vec{n}}$ where $\vec{n}$ are the normal vectors used in the engineering description. Again excluding the $\vec{[\overline{\sigma_p}]}=\vec{0}$ case the possibilities are:
\begin{itemize}
\item The principal stresses are non zero and have the same sign. Equation $\scal{\vec{n}}{\vec{[\overline{\sigma_p}]}\vec{n}}=\vec{0}$ has no real solutions, the base of the stress quadric is empty and the quadric is only the the Latin plane.
\item The principal stresses are non zero and one differs in sign to the other two. The normals $\vec{n}$ trace a cone in $\mathbb{R}^3$ and the ideal lines in $\mathcal{P}^3$ are tangents to an ellipse in the ideal plane. These ideal lines are mapped to an ellipse in the Greek plane of ideal lines, which will be the base of the stress-quadric.
\item One of the the principal stresses is $0$, the other two have the same sign. The solution of $\scal{\vec{n}}{\vec{[\overline{\sigma_p}]}\vec{n}}=\vec{0}$ corresponds to a single point in the Greek plane of ideal lines, which point will be the base of the stress-quadric.
\item One of the the principal stresses is $0$, the other two differ in sign. The solution of $\scal{\vec{n}}{\vec{[\overline{\sigma_p}]}\vec{n}}=\vec{0}$ corresponds to a pair of intersecting lines in the Greek plane of ideal lines; forming the base of the stress quadric.
\item Two of the the principal stresses is $0$. The solution of $\scal{\vec{n}}{\vec{[\overline{\sigma_p}]}\vec{n}}=\vec{0}$ corresponds to (double) line in the Greek plane of ideal lines; forming the base of the stress quadric.
\end{itemize}

For the strain tensor $d\vec{\phi}$ has to lie on the reference plane $\psi_{ref}$, the Greek plane corresponding to lines of $\psi_{ref}$ will be the apex of the strain-conic. The base will lie in the Latin plane corresponding to all the lines passing through $p$ and we may reduce the problem to finding the zeros of  $\scal{\vec{u}}{\vec{[\overline{\epsilon_p}]}\vec{u}}$ where $\vec{u}$ will give the directions of the lines passing through $p$. Excluding the $\vec{[\overline{\epsilon_p}]}=\vec{0}$ case the possibilities of the strain conic are:

\begin{itemize}
\item The principal strains are non zero and have the same sign. Equation $\scal{\vec{u}}{\vec{[\overline{\epsilon_p}]}\vec{u}}=\vec{0}$ has no real solutions, the base of the strain quadric is empty and the quadric is only the the Greek plane.
\item The principal strains are non zero and one differs in sign to the other two. Vectors $\vec{u}$ trace a cone in $\mathbb{R}^3$ and the lines corresponding to these directions appear as an ellipse in the Greek plane, which will be the base of the strain-quadric.
\item One of the the principal strains is $0$, the other two have the same sign. The solution of $\scal{\vec{u}}{\vec{[\overline{\epsilon_p}]}\vec{u}}=\vec{0}$ corresponds to a single point in the Latin plane, which point will be the base of the strain-quadric.
\item One of the the principal strains is $0$, the other two differ in sign. The solution of $\scal{\vec{u}}{\vec{[\overline{\epsilon_p}]}\vec{u}}=\vec{0}$ corresponds to a pair of intersecting lines in the Latin plane; forming the base of the stress quadric.
\item Two of the the principal strains is $0$. The solution of $\scal{\vec{u}}{\vec{[\overline{\epsilon_p}]}\vec{u}}=\vec{0}$ corresponds to (double) line in the Latin plane; forming the base of the strain quadric.
\end{itemize}

\section{Material laws}
As projective geometry has a duality principle to which we kept our description so far, it is natural to ask if the material law could be described as a correlation (dual-transformation). Especially knowing that in the case of elastic rods the ellipse of elasticity is a geometric representation of such correlation. Unfortunately the answer is negative: The material law is not representable with a correlation; even in the planar, linear, homogeneous, isotropic case. To see this consider point $p$ having stress state $\sigma_{12}=1$ with $\sigma_{11}=\sigma_{22}=0$. The strain-state is $\epsilon_{12}=2(1+\nu)/Y$ with $\epsilon_{11}=\epsilon_{22}=0$ where $\nu$ denotes Poisson's ratio and $Y$ Young's modulus. The stress conic is the pair of lines $\chi_1$ and $\chi_2$ passing through $p$ in the $e_1$ and $e_2$ direction. The strain conic in this case consists of two points and a line on them,  the line is reference line $\psi_{ref}$ and the two points are $\delta_1=\chi_1 \cap \psi_{ref}$ and $\delta_2=\chi_2 \cap \psi_{ref}$. If there exists a correlation describing the behaviour it needs to map lines $\{\chi_1, \chi_2 \}$ to points $\{ \delta_1, \delta_2 \}$ in some pairing. Now consider the stress state $\sigma_{11}=1$ with $\sigma_{22}=0=\sigma_{12}$. The strain state is $\epsilon_{11}=1/Y$, $\epsilon_{22}=-\nu/Y$ with $\epsilon_{12}=0$. The stress-conic is the line $\chi_2$. If there exists a correlation representing the material-law, the strain-conic needs to be either $\delta_1$ or $\delta_2$. As $\epsilon_{11} \neq 0$ and $\epsilon_{22} \neq 0$, neither is a good choice, meaning we can not represent the material law as a correlation.

\section{2.5 dimensional problems}
It is sometimes useful to think of the Airy stress function of planar problems through its graph, a surface in $\mathcal{P}^3$, while the spatial problems of some membrane shells may be analyzed by projecting the problem to a plane orthogonal to the load on the shell. The problems are even connected by Pucher's equation. Here we will provide some results on the projective transformations of these problems.

\subsection{Transforming the graph of the Airy stress function}
Let us embed our planar problem into $\mathcal{P}^3$ such that the plane of the problem corresponds to the projective plane $\vec{\alpha}_3(\vec{p})=0$. The Airy stress function is known \cite{phillips} to correspond to moments as follows: Let us denote the stress function with $\Psi(p)$ and consider a planar curve $p(s)$ running in the material. Denoting the resultant collected from $s=0$ acting on the material on the left side of the curve (according to the direction given by the parametrization $s$) with $\vec{f}(s)$, the value of the Airy stress function is the moment of the resultant with respect to the point on the curve (its work on a unit rotation orthogonal to the plane):
\begin{align}
\Psi(p(s))=\dag\left(\vec{e_3}\wedge \vec{p}\right) \left( \vec{f}(s)\right)=\star\left(\vec{e_3}\wedge \vec{p} \wedge \vec{f}(s)\right)
\end{align}    
for all $s$. We are interested in projective transformations mapping the planar problem into another one in the same plane, leaving the plane of the problem invariant. Furthermore, we wish to graph the Airy stress functions of both problems in the $e_3$ direction, meaning the ideal point in the $e_3$ direction needs to be invariant. The matrix describing such transformation will be of shape
\begin{align}
\vec{T}:=
\begin{bmatrix}
\vec{B} & \vec{0}_{3 \times 1} \\
\vec{0}_{1 \times 3} & \lambda_3
\end{bmatrix} \label{eq:inducedmap}
\end{align}
where $\lambda_3 \in \mathbb{R} \setminus \{0\}$ and the block $\vec{B}$ is invertible. It turns out if we are given a planar transformation with the corresponding $\vec{B}$ and $\lambda_f$ we can create an induced map $\vec{B}^I$ by setting $\lambda_3=|\vec{B}|\lambda_f$ in \eqref{eq:inducedmap} that will satisfy:
\begin{prop}
The induced map $\vec{B}^I$ maps the graph of the Airy stress function to the graph of the transformed Airy stress function.
\end{prop}
\begin{proof}
As the stress function is defined on the whole plane we omit the arclength parametrization of the curve from the notation. To see how the stress function transforms we may write according to Equation \eqref{eq:worktrafo}: 
\begin{align}
 |\vec{B}^I| \Psi(p)=\star\left(\vec{B}^I\vec{e_3}\wedge \vec{B}^I\vec{p} \wedge C_2 \vec{B}^I \vec{f}\right)\\
| \vec{B}|^2 \lambda_f \Psi(p)=\star\left(|\vec{B}|\lambda_f\vec{e_3} \wedge \lambda_p \vec{p}' \wedge C_2 \vec{T} \vec{f}\right)
\end{align}
which together with the mechanical interpretation of the stress function implies
\begin{align}
\Psi'(p')=\lambda_p^{-1} \lambda_f |\vec{T}| \Psi(p).
\end{align}
The graph of the stress function consists of points having representants
\begin{align}
\left(1,\vec{\alpha}_1(\vec{p}),\vec{\alpha}_2(\vec{p}),\Psi(p)\right)^T
\end{align}
which transform into
\begin{align}
\left(\lambda_p,\vec{\alpha}_1(\vec{B}^I\vec{p}),\vec{\alpha}_2(\vec{B}^I\vec{p}),|\vec{B}|\lambda_f\Psi(p)\right)^T \sim \left(1,\vec{\alpha}_1(\vec{p}'),\vec{\alpha}_2(\vec{p}'),\Psi'(p')\right)^T
\end{align}
completing the proof.
\end{proof}
 
\subsection{Membrane shells}
To pair our previous planar problem with membrane shells suppose we are given a membrane shell where the shape of the shell may be expressed as 
\begin{align}
\left(1,\vec{\alpha}_1(\vec{p}),\vec{\alpha}_2(\vec{p}),H(p) \right)^T
\end{align}
which is the graph of the height function $H(p)$. If the shell is subjected to loading in the $e_3$ direction with intensity $c(p)$, Puchers' equation \cite{csonka} gives a correspondence between functions $c(p)$, $H(p)$ and the Airy stress function $\Psi(p)$ of the continuum problem that is the projection of the membrane problem to the plane orthogonal to the $e_3$ direction. Due to the way the induced maps $\vec{B}^I$ are defined, they will commute with projection in the $e_3$ direction which we omit to show since it is technical and straightforward. The difficulty in the use of this method in the analysis of membrane shells is expected to come from the fact how the load function transforms. Assuming that the load corresponds to a stress state where the only non-zero component is the $e_3$ directional normal stress with the intensity of the load, the transformation rule may be expressed from Equation \eqref{eq:fesztrafo} as
\begin{align}
c'(p')=\lambda_p ^4 c(p)
\end{align}      
(we used the Symbolic Toolbox of MatLab). 

\section{Example application to tapered beams}
As an example, we look at the problem of tapered beams that have thin rectangular cross sections and only the top and bottom of the beams are subjected to taper. This implies we can treat them as a planar problems. The idea generalizes to rods where the sides are also tapered, requiring a $3$ dimensional treatment. We picked this problem to demonstrate what we derived, on a problem that is investigated today. The fact that applying the classic formulas of the Euler-Bernoulli beam theory to the trapezoidal problem gives erroneous results and the error grows with the growth of the taper angle has been long-since known \cite{10.1115/1.3636564}. Yet, there are contemporary works \cite{SU2022849,BERTOLINI2019527} arguing that for small taper angles and with some additional considerations results assuming the Bernoulli-Navier Hypothesis are good enough approximations. We laid this out to explain that we will give a strategy to find a statically possible stress distribution and this strategy is entirely insensitive to the taper angle. We do not claim that a tapered rod of linear isotropic material chooses the stress distribution our strategy finds, the transformation of the material properties seems to be a more complex question beyond the scope of the present paper.\\

\begin{figure}
\begin{center}
\includegraphics[width=0.75\textwidth]{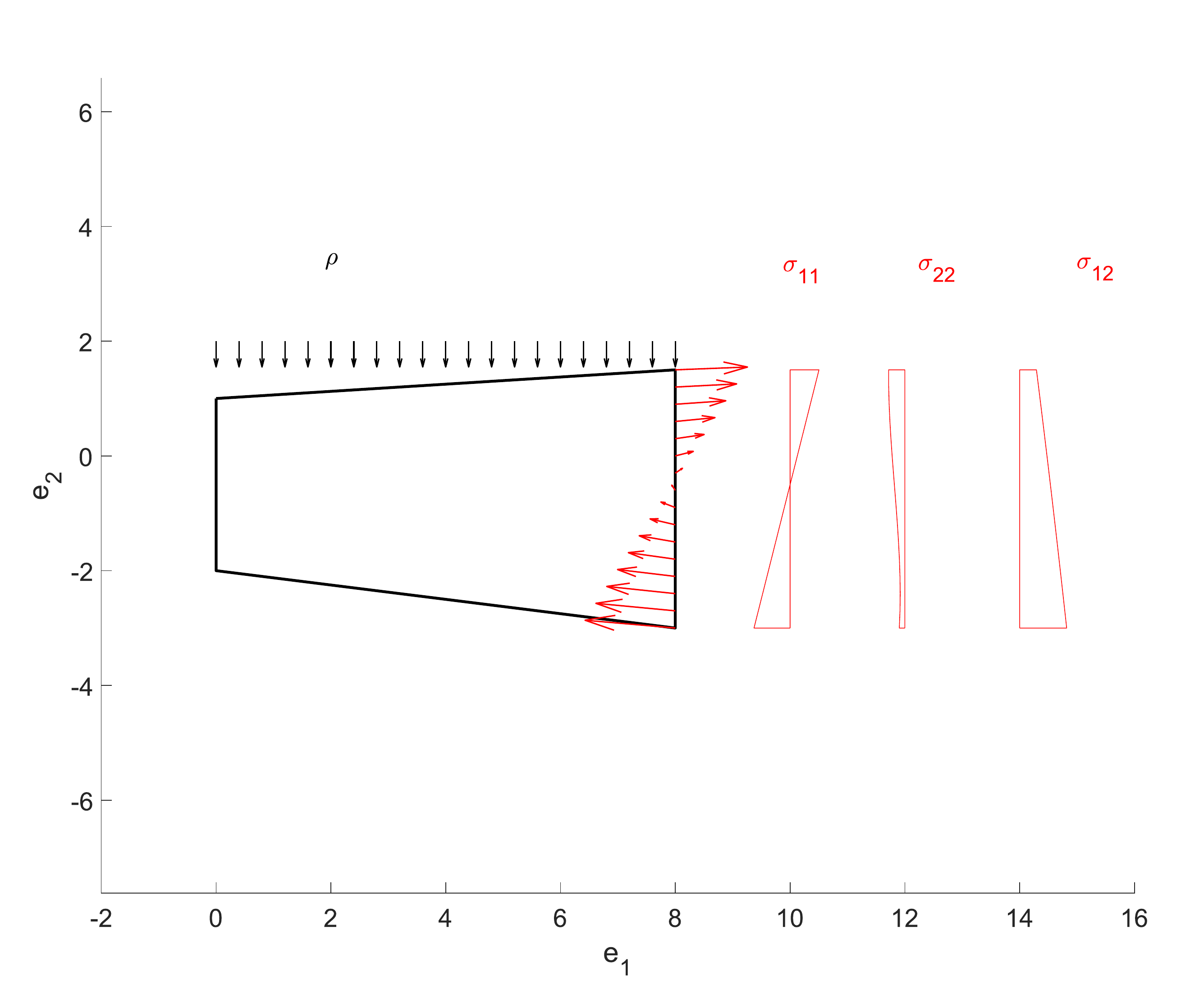}
\end{center}
\caption{Example tapered rod problem. The stresses have been rescaled for better readability. The arrows were drawn with proportions load:reaction = 10:1 while the stress diagrams with $\sigma_{11}:\sigma_{22}:\sigma_{12} = 1:15:15$.}\label{fig:e}
\end{figure}

\begin{figure}
\begin{center}
\includegraphics[width=0.75\textwidth]{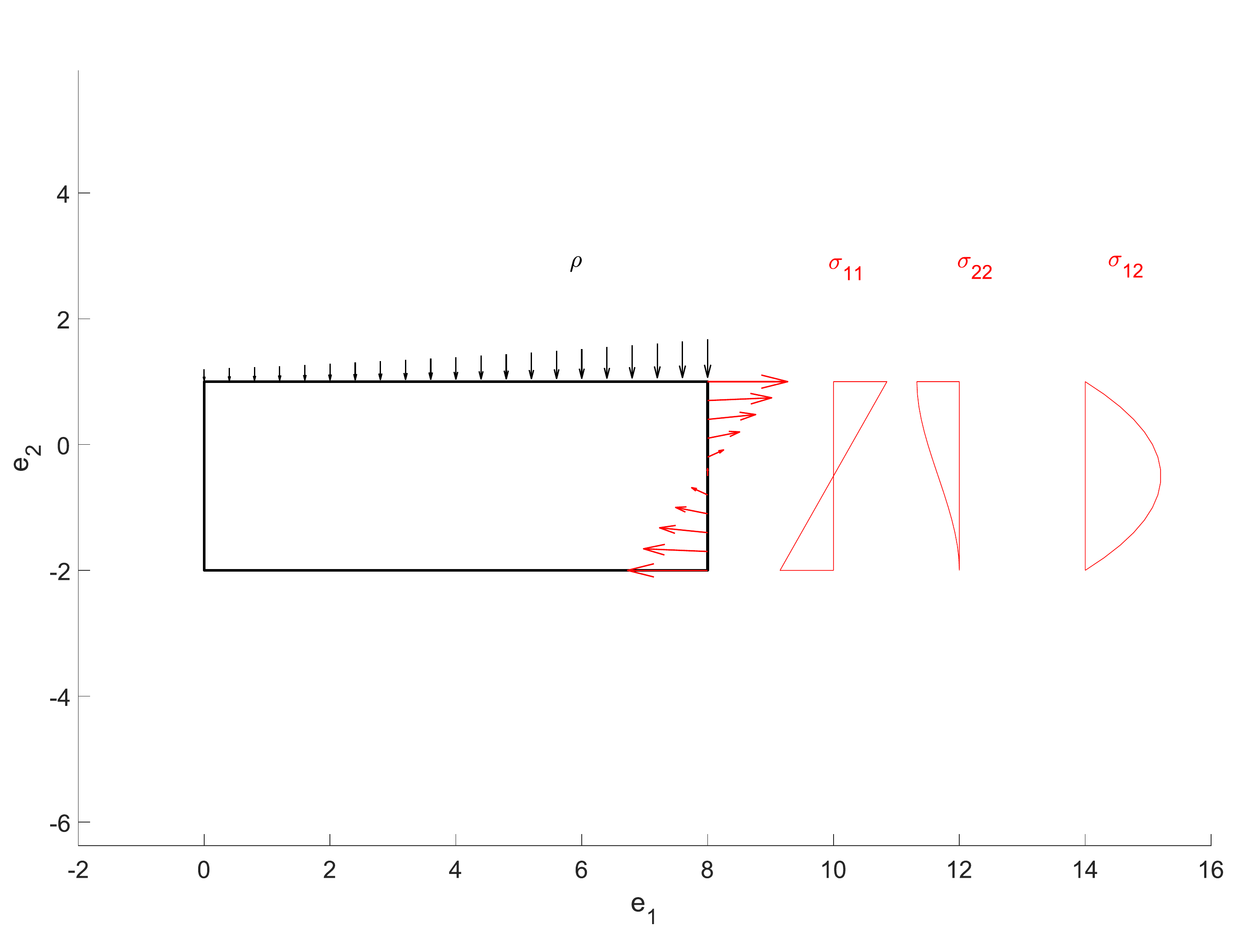}
\end{center}
\caption{The straight uniform rod the tapered example was transformed into. The stresses have been rescaled for better readability. The arrows were drawn with proportions load:reaction = 10:1 while the stress diagrams with $\sigma_{11}:\sigma_{22}:\sigma_{12} = 1:15:15$.}\label{fig:f}
\end{figure}

Consider a cantilever trapezoidal plate with corners $\vec{p}_1=(1,0,-h_2)$, $\vec{p}_2=(1,0,h_1)$, $\vec{p}_3=(1,L,H_1)$ and $\vec{p}_4=(1,L,-H_2)$ supported along the edge $p_3,p_4$. The load acts along the $p_2,p_3$ edge, is $-e_2$ directional with intensity $\rho$ (uniformly distributed along the $e_1$ axis). A numerical version of the problem is drawn in Figure \ref{fig:e} along with the reaction stresses. To transform the problem back to a rod of constant cross-section we may chose to have $\vec{p}'_1=\vec{p}_1$, $\vec{p}'_2=\vec{p}_2$, $\vec{p}'_3=(1,L,h_1)$ and $\vec{p}'_4=(1,L,-h_2)$.
The $4$ equations $\vec{p}'_i \sim \vec{T}\vec{p}_i$ give 8 equations that determine $\vec{T}$ up to a scalar multiple and it can be seen that
\begin{align}
\vec{T}=
\begin{bmatrix}
1 & (H_1 + H_2 - h_1 - h_2)/(L (h_1 + h_2)) & 0\\
0 & (H_1 + H_2)/(h_1 + h_2) & 0\\
0 & -(H_1 h_2 - H_2 h_1)/(L (h_1 + h_2))  & 1
\end{bmatrix}
\end{align}
is a good choice for $\vec{T}$. We may also note that $\vec{Tx}_2=\vec{x}_2$ implying all $x_2$ directional lines stay parallel with themselves. This will mean the load will stay parallel with the $e_2$ direction. The intensity of the transformed load may be determined by identifying the load with a stress state as $\sigma_{11}=0=\sigma_{12}$, $\sigma_{22}=\rho$. Using Equation \eqref{eq:fesztrafo} we have
\begin{align}
\rho'=\rho \left( \frac{x_1 (H_1 + H_2 - h_1 - h_2)}{L(h_1 + h_2)} +1 \right)^3 \frac{h_1+h_2}{H_1+H_2}
\end{align}
for the transformed problem. A numerical version of this transformed problem is drawn in Figure \ref{fig:f} along with the reaction stresses. Stress components $\sigma_{11}$ and $\sigma_{12}$ may be computed from the formulas from Euler-Bernoulli beam theory. After this $\sigma_{22}$ becomes computable from the $e_2$ directional equilibrium of stresses of the upper part of the cross-section according to how the Zhuravskii Shear Stress Formula partitions it. The solution may be then transformed back to the tapered beam. While the expressions of the stresses are closed form, we do not present them here as they are long and unfit for pen and paper use. The interested reader should be able to reconstruct them with a symbolic computation software.


\section{Acknowledgements}
The author is very grateful to dr. István Sajtos for the idea to showcase  the theory on tapered rods and to reduce the problem of finding a statically possible stress distribution to the Euler-Bernoulli beam theory.

\section{Declarations}
%\subsection{Funding}
This work was supported by the Ministry of Innovation and Technology of Hungary from the National Research, Development and Innovation Fund under the PD 142720 funding scheme.

\bibliographystyle{unsrt} 
\bibliography{latexbib}

\begin{thebibliography}{10}

\bibitem{plucker1868neue}
J.~Pl{\"u}cker.
\newblock {\em Neue Geometrie des Raumes: gegr{\"u}ndet auf die Betrachtungen
  der geraden Linie als Raumelement}.
\newblock Neue Geometrie des Raumes: gegr{\"u}ndet auf die Betrachtung der
  geraden Linie als Raumelement : zwei Abtheilungen in einem Bande. Teubner,
  Leipzig, 1868.

\bibitem{plucker_mechanics}
J.~Pl{\"u}cker.
\newblock Fundamental views regarding mechanics.
\newblock {\em Philosophical Transactions of the Royal Society of London},
  156:361--380, 1866.

\bibitem{davidson2004robots}
J.K. Davidson and K.H. Hunt.
\newblock {\em Robots and Screw Theory: Applications of Kinematics and Statics
  to Robotics}.
\newblock Oxford University Press, Oxford, 2004.

\bibitem{gallardo2016kinematic}
J.~Gallardo-Alvarado.
\newblock {\em Kinematic Analysis of Parallel Manipulators by Algebraic Screw
  Theory}.
\newblock Springer International Publishing, eBook, 2016.

\bibitem{ball1900treatise}
R.S. Ball.
\newblock {\em A Treatise on the Theory of Screws}.
\newblock Cambridge Mathematical Library. Cambridge University Press,
  Cambridge, 1900.

\bibitem{crapo1982statics}
Henry Crapo and Walter Whiteley.
\newblock Statics of frameworks and motions of panel structures: a projective
  geometric introduction.
\newblock {\em Structural Topology, 1982, num. 6}, 1982.

\bibitem{rankine1857transformations}
William John~Macquorn Rankine.
\newblock Ii. on the mathematical theory of the stability of earth-work and
  masonry.
\newblock {\em Proceedings of the Royal Society of London}, 8:60--61, 1857.

\bibitem{affinetransformations_2006_Ostrosablin}
N.~I. Ostrosablin.
\newblock Affine transformations of the equations of the linear theory of
  elasticity.
\newblock {\em Journal of Applied Mechanics and Technical Physics},
  47:564–572, 2006.

\bibitem{csonka}
P.~Csonka.
\newblock {\em Theory and Practice of Membrane Shells}.
\newblock VDI Verlag, Duesseldorf, 1987.

\bibitem{prasolov1994problems}
V.V. Prasolov and S.~Ivanov.
\newblock {\em Problems and Theorems in Linear Algebra}.
\newblock History of Mathematics. American Mathematical Society, 1994.

\bibitem{aitken2017determinants}
A.C. Aitken.
\newblock {\em Determinants and Matrices}.
\newblock Read Books Limited, 2017.

\bibitem{richter2011perspectives}
J.~Richter-Gebert.
\newblock {\em Perspectives on Projective Geometry: A Guided Tour Through Real
  and Complex Geometry}.
\newblock Springer, Berlin, 2011.

\bibitem{csikos-kiss}
Balázs Csikós and Görgy Kiss.
\newblock {\em Projetkív geometria}.
\newblock SZTE Bolyai Intézet, Szeged, 2011.

\bibitem{phillips}
H.~B. Phillips.
\newblock Stress functions.
\newblock {\em Journal of Mathematics and Physics}, 13(1-4):421--425, 1934.

\bibitem{10.1115/1.3636564}
Bruno~A. Boley.
\newblock {On the Accuracy of the Bernoulli-Euler Theory for Beams of Variable
  Section}.
\newblock {\em Journal of Applied Mechanics}, 30(3):373--378, 09 1963.

\bibitem{SU2022849}
Xiaolong Su and Man Zhou.
\newblock Analysis of shear stresses in tapered beams under bending, shear and
  axial force.
\newblock {\em Structures}, 41:849--865, 2022.

\bibitem{BERTOLINI2019527}
P.~Bertolini, M.A. Eder, L.~Taglialegne, and P.S. Valvo.
\newblock Stresses in constant tapered beams with thin-walled rectangular and
  circular cross sections.
\newblock {\em Thin-Walled Structures}, 137:527--540, 2019.

\end{thebibliography}

\end{document}